\documentclass{aa}
\usepackage[varg]{txfonts}
\usepackage{graphicx}
\usepackage{color}
\def\gsim{~\rlap{$>$}{\lower 1.0ex\hbox{$\sim$}}}

\begin{document}

\title{Enhanced direct collapse due to Lyman $\alpha$ feedback}


\author{Jarrett L. Johnson\inst{1}\thanks{\emph{Email address:} 
    jlj@lanl.gov}
  \and Mark Dijkstra\inst{2}}


\institute{X Theoretical Design and the Center for Theoretical Astrophysics, Los Alamos National Laboratory, Los
  Alamos, NM 87545, USA
  \and Institute of Theoretical Astrophysics, University of Oslo,
  P.O. Box 1029 Blindern, N-0315 Oslo, Norway} 



\abstract {We assess the impact of trapped Lyman $\alpha$ cooling
radiation on the formation of direct collapse black holes (DCBHs).  We apply a one-zone chemical and thermal evolution model,
accounting for the photodetachment of H$^{-}$ ions, precursors to the
key coolant H$_{\rm 2}$, by Lyman $\alpha$ photons produced during the
collapse of a cloud of primordial gas in an atomic cooling halo at high
redshift.  We find that photodetachment of H$^{-}$ by trapped Lyman $\alpha$ photons may lower the level of the
H$_{\rm 2}$-dissociating background radiation field required for DCBH
formation substantially, dropping the critical flux
by up to a factor of a few.  This translates into a potentially large
increase in the expected number density of DCBHs in the early
Universe, and supports the view that DCBHs may be the seeds for the BHs residing
in the centers of a significant fraction of galaxies today.  We
find that detachment of H$^-$ by Lyman $\alpha$ has the strongest
impact on the critical flux for the relatively high background radiation
temperatures expected to characterize the emission from young, hot
stars in the early Universe.  This lends support to the DCBH origin of
the highest redshift quasars.}

\keywords{radiative transfer -- cosmology:  theory -- black hole physics --
 dark ages, reionization, first stars -- quasars:  supermassive black holes -- molecular processes}
\maketitle

\section{Introduction}
The direct collapse scenario for black hole (BH) formation in the
early Universe has received much attention in recent years, in
particular for its ability to explain the formation of BHs
with masses $\ga$ 10$^9$ M$_{\odot}$ within the first billion years of
cosmic history (e.g. Mortlock et al. 2011; Wu et al. 2015).  The key
ingredients for the formation of the massive ($\sim$ 10$^5$
M$_{\odot}$) seed BHs in this theory are (1) primordial gas collapsing
into an atomic cooling dark matter halo and (2) a sufficiently low fraction of
H$_{\rm 2}$ molecules in the gas to prevent cooling below the $\sim$
10$^4$ K cooling limit of atomic hydrogen (for reviews see Volonteri
2012; Haiman 2013; Johnson \& Haardt 2016; Latif \& Ferrara 2016).

The main ways that are envisioned for keeping the primordial gas devoid
of molecules is photodissociation of H$_{\rm 2}$ due to
so-called Lyman-Werner (LW) radiation at energies 11.2 - 13.6 eV and
photodetachment of the H$^{-}$ ion, which is an intermediary in the
formation of H$_{\rm 2}$ (e.g. Bromm \& Larson 2004), by photons with
energies $>$ 0.76 eV (e.g. Chuzhoy et al. 2007).  The relative
importance of each of these processes has been found to be strongly
dependent on the spectrum of the incident radiation (e.g. Shang et al. 2010; 
Sugimura et al. 2014; Glover 2015; Agarwal et al. 2015; Latif et
al. 2015; Wolcott-Green et al. 2016), presumably
produced by a nearby star-forming galaxy (e.g. Dijkstra et al. 2008; 
Agarwal et al. 2012; Visbal et al. 2014; Regan et al. 2016a).  

An additional source of radiation which contributes to the
photodetachment of H$^{-}$ and so limits the formation rate of H$_{\rm
  2}$ is the trapped Lyman $\alpha$ cooling radiation that is emitted from the collapsing
atomic gas in the halo itself (Spaans \& Silk 2006; Schleicher et al. 2010).  
Here we explore the impact that this trapped radiation has on the production
of H$_{\rm 2}$ molecules in the gas and, in turn, on its ability to
cool below the $\sim$ 10$^4$ K required for direct collapse black hole
(DCBH) formation.  In the next Section, we outline the one-zone
chemical and cooling model that we employ for our study and we describe
our approach to modeling the photodetachment of H$^{-}$ by Lyman
$\alpha$ cooling radiation.   In Section 3 we present the basic results of
our calculations, and in Section 4 we explore the impact of Lyman
$\alpha$ feedback on the critical LW flux required for DCBH
formation.  Finally, we give our conclusions and provide a
brief discussion of our results in Section 5.

\section{Feedback from Lyman $\alpha$ Cooling Radiation}
For our study, we begin with the same one-zone model for the collapse 
of the primordial gas as presented in Johnson \& Bromm (2006), which
is very similar to other one-zone models that have been routinely 
applied to DCBH formation (e.g. Omukai et al. 2005, 2008; Schleicher
et al. 2010; Agarwal et al. 2016a).  The model assumes that the
density of the primordial gas increases on the free-fall timescale,
and the non-equilibrium chemical and thermal evolution of the
collapsing gas is calculated.  All of the pertinent primordial
chemical species are included, as are all of the pertinent radiative
processes.  

While the reader is referred to Johnson \& Bromm (2006) for more
details, here we describe the key ingredients in the model that we draw
on for our study of the direct collapse scenario.  One important update to this code has been the
  adoption of the H$_{\rm 2}$ self-shielding prescription presented in
  Wolcott-Green et al. (2011; see also Hartwig et al. 2015,
  Wolcott-Green et al. 2016), which replaced the simpler prescription
  presented in Bromm \& Loeb (2003).  We have also updated the
  collisional dissociation rate of H$_{\rm 2}$ to that presented in
  Martin et al. (1996), which is now the commonly adopted rate
  (e.g. Shang et al. 2010; Agarwal et al. 2016a).  The model includes the main cooling
  processes that are relevant for the direct collapse scenario, which are 
atomic hydrogen line cooling and molecular (H$_{\rm 2}$) line
cooling (e.g. Cen 1992; Abel et al. 1997).  In addition, the model
tracks the non-equilibrium chemistry of the primordial gas and the
formation of H$_{\rm 2}$ molecules, the main channel for
which is the following two reactions:

\begin{equation}
{\rm H} + {\rm e^-} \to {\rm H^-} + {\rm \gamma}    
\end{equation}

\begin{equation}
{\rm H^-} + {\rm H} \to {\rm H}_{\rm 2} + {\rm e^-} \mbox{\ ,}
\end{equation}
where e$^-$ is a free electron and {\rm $\gamma$} is a photon.  Given
that H$^-$ is the main precursor to H$_{\rm 2}$, the photodetachment of
H$^-$ is a key reaction to track in order to accurately calculate the
formation rate of H$_{\rm 2}$.  Thus, we track the photodetachment of
H$^-$ as well as the photodissociation of H$_{\rm 2}$ in our model,
adopting the rates presented in Shang et al. (2010) as functions of
the temperature of the radiation field.\footnote{While H$_{\rm 2}^+$ is
  also a precursor to H$_{\rm 2}$ formation in the primordial gas, the
rate of H$_{\rm 2}$ formation via this channel is much lower than that
through the H$^-$ channel for the relatively hot radiation spectra
($\ge$ 10$^4$ K) that are of interest here (see e.g. Sugimura et al. 2015).  For this
reason, we neglect the radiative destruction of H$_{\rm 2}^+$ in our modeling.}

We solve additional equations in order to assess the
impact of photodetachment of H$^{-}$ by Lyman $\alpha$ photons.   
To begin, we make the simple assumption that the luminosity of Lyman $\alpha$ cooling emission in the 
cloud balances the rate of gravitational potential energy release
during the collapse of the cloud (e.g. Dijkstra et al. 2016).  This is
a sound approximation, as it has been shown in numerous cosmological
simulations that the collapse of primordial gas in atomic cooling halos
is roughly isothermal and occurs on the free-fall timescale (e.g. Wise
et al. 2008; Regan \& Haehnelt 2009).  
Thus, we adopt the following simple expression for the Lyman $\alpha$ luminosity:

\begin{equation}
L_{\rm Ly \alpha} = \frac{G M_{\rm cloud}^2}{r_{\rm cloud}} \frac{1}{t_{\rm ff}}\mbox{\ ,}
\end{equation}
where $t_{\rm ff}$ = (3$\pi$ / 32$G$$\rho$)$^{\frac{1}{2}}$ is
the free-fall time, where $G$ is Newton's constant and $\rho$ is the
density of the collapsing gas.   Here $M_{\rm cloud}$ = 10$^6$ M$_{\odot}$ is the
typical mass of the central gas cloud collapsing in an atomic cooling
halo (e.g. Wise et al. 2008; Johnson et al. 2011, 2014; Latif et al. 2013; Choi
et al. 2013).  Assuming a uniform cloud density, which is appropriate
for our simplified one-zone calculations, this implies a cloud radius of
$r_{\rm cloud}=30$ ${\rm pc}$ $(n/10^2$ ${\rm cm}^{-3})^{-1/3}$
where $n$ is the number density of hydrogen nuclei.  As the gas cools, this is the characteristic length scale over which Lyman $\alpha$
photons must diffuse in order to escape the cloud.  

The diffusion of Ly$\alpha$ photons out of the cloud enhances the energy density in Ly$\alpha$ photons by an amount that depends on the cloud column density, $N_{\rm H}$. The total line center optical depth to Ly$\alpha$ is given by
$\tau_{\rm Ly \alpha} = 5.9 \times 10^6 \left(\frac{N_{\rm H}}{10^{20}
    \, {\rm cm^{-2}}}\right) \left(\frac{T}{10^4 \, {\rm      K}}
\right)^{-\frac{1}{2}}$,
where $T$ is the temperature of the gas and the column density of 
hydrogen atoms is $N_{\rm H}$ = $r_{\rm cloud} n$ (e.g. Osterbrock \&
Ferland 2006).  Following Adams
(1975, see also Smith et al. 2017 for an updated discussion), the pathlength traversed by the photons in escaping
the cloud is enhanced by a factor $\mathcal{M}_{\rm F}\sim  (a_{\rm v} \tau_{\rm Ly
  \alpha}$)$^{\frac{1}{3}}$, where $a_{\rm v}$ = 4.7 $\times$10$^{-4}$
($T$ / 10$^4$ K)$^{-\frac{1}{2}}$ is the Voigt parameter.  We estimate the total energy density
  in Ly$\alpha$ radiation, $u_{\alpha}$, to be
  \begin{equation}\label{eq:ua}
  u_{\alpha}=\mathcal{M}_{\rm F}\frac{L_{\rm Ly \alpha} r_{\rm cloud}}{V_{\rm cloud}c},
  \end{equation} where $V_{\rm cloud}$ denotes the volume of the cloud, and $\mathcal{M}_{\rm F}$ equals
\begin{equation}\label{eq:mf}
\mathcal{M}_{\rm F}\sim 60\left(\frac{n}{10^2\hspace{1mm}{\rm
      cm}^{-3}}\right)^{2/9}\left(
  \frac{M}{10^6M_{\odot}}\right)^{1/9} \left(\frac{T}{10^4 \, {\rm      K}}\right)^{-1/3}.
\end{equation}
Assuming that the Lyman $\alpha$ radiation field is isotropic
within the cloud due to the large optical depth to scattering, we can then
approximate the photodetachment rate $R_{\rm detach}$ of H$^{-}$ ions
by Lyman $\alpha$ photons as 

\begin{equation}\label{eq:detach}
R_{\rm detach} = \sigma_{\rm H-} \mathcal{M}_{\rm F}\frac{L_{\rm Ly
    \alpha}}{E_{\rm Ly \alpha}} \frac{3}{4 \pi r_{\rm cloud}^2}\mathcal{B}   \mbox{\ ,}
\end{equation}
where the cross section for this process
is $\sigma_{\rm H-}$ = 5.9 $\times$ 10$^{-18}$ cm$^2$ and $E_{\rm Ly \alpha}$ = 10.2 eV is the
energy of a Lyman $\alpha$ photon (e.g. de Jong 1972; Shapiro \& Kang
1987).  As the dependence on the optical depth to scattering shows, this rate is
elevated due to the many scatterings that Lyman
$\alpha$ photons make in passing out of the cloud.  Finally,
$\mathcal{B}$ accounts for the fact that spatial diffusion of Ly$\alpha$ photons
does not necessarily uniformly enhance the Ly$\alpha$ intensity
throughout the cloud, especially when Ly$\alpha$ emission is
concentrated more towards the center of the cloud (see Fig. A.1). In
the Appendix we show that $\mathcal{B}$ can be as large as $\mathcal{B} \sim 10$ toward the center of the cloud, which is where the DCBH forms. Throughout, we will investigate the impact of varying $\mathcal{B}$ within the range $\mathcal{B}=1-10$.

Combining the above equations, we obtain the following expression for the
photodetachment rate as a function of cloud temperature, mass and
density:

\begin{equation}
R_{\rm detach} \simeq 10^{-8} \, {\rm s}^{-1} \left(\frac{M_{\rm cloud}}{10^6 \, {\rm M}_{\odot}}
\right)^{\frac{10}{9}} \left(\frac{T}{10^4 \, {\rm K}}  \right)^{-\frac{1}{3}}
\left(\frac{n}{10^{2} \, {\rm cm}^{-3}}   \right)^{\frac{31}{18}} \left(\frac{\mathcal{B}}{2}\right) \mbox{\ .}
\end{equation}
This is the equation that we include in our calculations in order to
assess the role that Lyman $\alpha$ feedback plays in the formation of
DCBHs.

\section{Basic Results}
Here we show our results for two sets of calculations, one in which
the effect of H$^{-}$ detachment by Lyman $\alpha$ photons is included
and another in which it is neglected.  In both cases, we also include
the effect of a background LW radiation field, which is
assumed to only contribute to the photodissociation of H$_{\rm 2}$
molecules and not to the detachment
of H$^{-}$ ions.  In the next Section, we explore how the inclusion of the
photodetachment rate due to the background radiation impacts the
evolution of the collapsing gas.  Finally, here we only consider cases with
$\mathcal{B}$ = 1, corresponding to the simplest case of uniform Lyman
$\alpha$ emission from the collapsing cloud.  We explore cases with
higher $\mathcal{B}$ values, corresponding to strongly centralized
emission, in the next Section.

\begin{figure}
  \includegraphics[angle=0,width=3.4in]{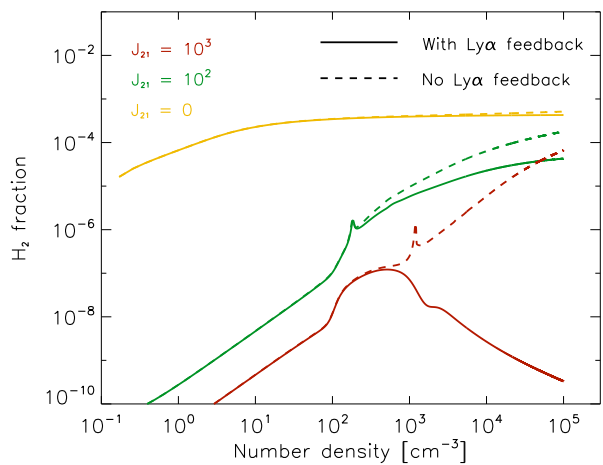}
  \caption{The evolution of the H$_{\rm 2}$ molecule fraction as a function of the number density of hydrogen
    nuclei, with ({\it solid lines}) and without ({\it dashed lines}) accounting for the effect of
    photodetachment of H$^{-}$ ions by Ly$\alpha$ photons.  The colors
    denote calculations assuming different background radiation
    fields, as labeled,
    which are assumed only to dissociate H$_{\rm 2}$ molecules.  In
    all cases shown here $\mathcal{B}$ = 1, corresponding to uniform
    Lyman $\alpha$ emission within the collapsing cloud.}
\end{figure}

\begin{figure}
  \includegraphics[angle=0,width=3.4in]{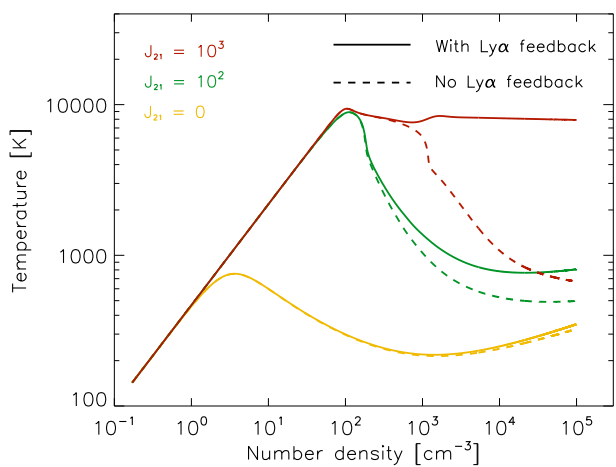}
  \caption{The evolution of the gas
    temperature as a function of the number density of hydrogen
    nuclei, with ({\it solid lines}) and without ({\it dashed lines}) accounting for the effect of
    photodetachment of H$^{-}$ ions by Ly$\alpha$ photons.  The colors
    denote calculations assuming different background radiation fields
    which are assumed only to dissociate H$_{\rm 2}$ molecules.  With no photodetachment the temperatures remain too low
    for DCBH formation in all cases, but with this effect included DCBH formation can
    occur for a background radiation field with $J_{\rm 21}$ $\sim$
    10$^3$.  As in Figure 1, here $\mathcal{B}$ = 1, corresponding to uniform
    Lyman $\alpha$ emission within the collapsing cloud.}
\end{figure}

Figure 1 shows the evolution of the H$_{\rm 2}$ fraction of the gas,
as a function of density, both with and without the above equations
for Lyman $\alpha$ photodetachment included.  The three sets of curves
correspond to different values of the LW background
radiation field $J_{\rm 21}$, which is expressed in the standard units
of 10$^{-21}$ erg s$^{-1}$ cm$^{-2}$ Hz$^{-1}$ sr$^{-1}$.  As expected,
the H$_{\rm 2}$ fraction is steadily depressed as the level of the
background radiation increases.  The impact of Lyman $\alpha$
photodetachment is also evident, resulting in the peak H$_{\rm 2}$
abundances dropping by orders of magnitude in the cases with
relatively high background radiation levels $J_{\rm 21}$ $>$ 100.  

The thermal evolution of the gas in these same sets of calculations is
shown in Figure 2.\footnote{Note that we recover the canonical cooling behavior of the gas for
the case with no background radiation ($J_{\rm 21}$ = 0) and no
H$^{-}$ photodetachment (e.g. Bromm \& Larson 2004; Greif et
al. 2015), as expected since we are employing effectively the same
code as in previous studies of such processes (Johnson \& Bromm
2006).}    Due to the depressed H$_{\rm 2}$ fraction, molecular cooling is less
effective with higher levels of the background radiation field.
However, in all cases shown here the gas is still able to cool to $\la$
10$^3$ K when H$^{-}$ detachment is not included in the calculation.
With this effect included, the cooling of the gas is suppressed at high density, resulting in much higher
temperatures.  Importantly, we find that with photodetachment
included, the temperature remains high enough for DCBH formation in
the case with $J_{\rm 21}$ = 10$^3$.  Thus, H$^-$ detachment by Lyman
$\alpha$ photons has the effect of lowering the critical value of the background
radiation level required for the formation of DCBHs.

\begin{figure}
  \includegraphics[angle=0,width=3.4in]{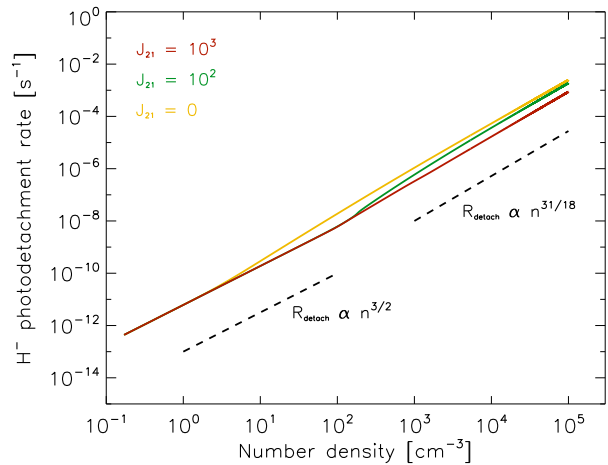}
  \caption{The rate of H$^{-}$ photodetachment by Ly$\alpha$ cooling
    radiation (equation 7), for the same calculations shown in Figures
    1 and 2.   The
    photodetachment rate is slightly higher for lower levels of the background
    radiation $J_{\rm 21}$ due to the temperature dependence of the
    cross section for Lyman $\alpha$ scattering.  At low
    densities the gas evolves adiabatically, leading to the scaling
    $R_{\rm detach}$ $\propto$ $n^{3/2}$, whereas at higher densities
    the scaling is better approximated assuming the gas is isothermal,
    leading to the scaling $R_{\rm detach}$ $\propto$ $n^{31/18}$.  Note
    that, in all cases, the photodetachment rate rises above the
    critical rate of $\sim$ 10$^{-5}$ s$^{-1}$ found in calculations assuming a constant background
    rate and a weak H$_{\rm 2}$-dissociating radiation field by Agarwal et al. (2016a).}
\end{figure}

To more fully elucidate the impact of photodetachment, the photodetachment rates in our calculations, as a function
of the cloud density, are shown in Figure 3.  The density and
time dependence of the photodetachment rate makes comparison
with previous determinations of the critical rate of photodetachment for
DCBH formation difficult (e.g. Sugimura et al. 2014; Agarwal et
al. 2016a; Wolcott-Green et al. 2016), as constant photodetachment
rates have typically been assumed.  However, it is clear that the photodetachment rates we
find rise well above the critical value of $\sim$ 10$^{-5}$ s$^{-1}$ found, for instance, by Agarwal et
al. (2016a) for the case of a weak H$_{\rm 2}$-dissociating radiation field.  Thus, in this sense, our results are consistent with, and
can be understood in the context of, previous work.  Noting from
Figure 2 that the
gas evolves roughly adiabatically up to $n$ $\sim$ 10$^2$ cm$^{-3}$ such that $T$ $\propto$ $n^{2/3}$, the scaling $R_{\rm detach}$ $\propto$
$n^{3/2}$ provides a good match to our calculations, as shown in
Figure 3.  At higher densities an isothermal scaling of $R_{\rm detach}$ $\propto$
$n^{31/18}$ provides a better fit, as is also shown in the Figure.  We next turn to
assessing the impact of Lyman $\alpha$ feedback on the value of the
critical LW flux required for DCBH formation.

\begin{table*}
\centering                                      
\begin{tabular}{c c c c}          
\hline\hline                        
\\
$T_{\rm rad}$ [K] & Ly$\alpha$ feedback ($\mathcal{B}$  = 1) & Ly$\alpha$ feedback ($\mathcal{B}$  = 10)  & No Ly$\alpha$ feedback \\    
\\
\hline                                   
\\ 
   10$^4$  & 24   & 22 & 26 \\      
    10$^5$  & 900 &  200  & 1100       \\
\\
\hline                                             
\\
\end{tabular}
\caption{Our calculated values of the critical LW radiation field $J_{\rm
    21,crit}$ required for DCBH formation, for two different background
  radiation temperatures, with and without accounting for the impact
  of photodetachment of H$^-$ by Lyman $\alpha$ cooling radiation,
  which is assumed to be either uniform ($\mathcal{B}$ = 1) or
  strongly centrally concentrated ($\mathcal{B}$ = 10).
  While the photodetachment rate due to the background radiation is
  relatively high already when $T_{\rm rad}$ is relatively low, the detachment
  rate due to Lyman $\alpha$ feedback comes to dominate the rate due to the
  background when $T_{\rm rad}$ is high, in particular at high gas
  densities (see Figure 3).}              
\label{table:1}      
\end{table*}

\section{The Impact on the Critical Lyman-Werner Flux}
Here we consider how our results change when including the H$^-$
photodetachment rate due to the background radiation.  To do so, we
carry out the same calculations as shown in Figure 2, but now
including also the H$^-$ detachment rate due to the background
radiation field.  We adopt the rates presented in Shang et al. (2010) assuming simple
blackbody spectra at $T_{\rm rad}$ = 10$^4$ and 10$^5$ K, and we
evaluate the critical LW flux $J_{\rm 21,crit}$ that is
required to maintain the gas at $\sim$ 10$^4$ K, leading to the
formation of a DCBH.  

Our results are presented in Figures 4 and 5, and are summarized in Table 1.
As shown in the left panels of Figures 4 and 5,
for a relatively low background radiation temperature of $T_{\rm rad}$
= 10$^4$ K 
the additional suppression of H$_{\rm 2}$ cooling due to Lyman
$\alpha$ feedback is relatively small, as the LW flux required to
maintain the gas at $\simeq$ 10$^4$ K at a density of $n$ $\sim$ 10$^5$ cm$^{-3}$ is $J_{\rm 21,crit}$ $\simeq$ 26 neglecting the
effect and $J_{\rm 21,crit}$ $\simeq$ 22 - 24, depending on the
geometry of the Lyman $\alpha$ emission (i.e. for $\mathcal{B}$ = 1 - 10), when accounting for it.  However,
as shown in the right panels of Figures 4 and 5, for a larger background radiation
temperature of $T_{\rm rad}$ = 10$^5$ K accounting for Lyman $\alpha$
feedback results in a much larger drop in the critical flux from $J_{\rm 21,crit}$
$\simeq$ 1.1 $\times$ 10$^3$ to $\simeq$ 200 - 900, depending on the
geometry of the Lyman $\alpha$ emission.  Thus, for the spectra
expected from hot, young stars in the early universe, the impact of
Lyman $\alpha$ feedback may be especially important.  As shown in
Figure 3 and in equation (7), it is detachment rates due to Lyman
$\alpha$ feedback at high densities, which are higher than the detachment rate
due to the background radiation field, that result in a lower critical
LW background flux.  

It is important to note the reason for the much larger difference in
the critical flux in the case of the higher background radiation
temperature.  This is ultimately due to the much lower rate of
H$^-$ photodetachment, relative to the H$_{\rm 2}$ photodissociation
rate, for the higher temperature background radiation field.
Specifically, the photodetachment rate at a given value of $J_{\rm
  21}$ is some four orders of
magnitude lower for a background temperature of 10$^5$ K than it is
for one of 10$^4$ K (Shang et al. 2010).  This implies that the 
rate of photodetachment by Lyman $\alpha$ photons,
which is independent of the spectrum of the background radiation
field, is much higher relative to the rate due to the background
radiation for the hotter spectrum than it is for the colder one.
This leads directly to the much larger drop in the critical value of
$J_{\rm 21}$ due to Lyman $\alpha$ feedback that we find for the
hotter background spectrum than for the colder one.

The values we find for the critical LW flux ($J_{\rm
  21,crit}$) in the cases neglecting Lyman $\alpha$ feedback 
are broadly consistent with the values found by previous authors
(see e.g. Omukai et al. 2008; Sugimura et al. 2014; Latif
et al. 2015; Hartwig et al. 2015; Agarwal et al. 2016a,b; Glover 2016), although they are 
different in detail due to differences in the models adopted in these
studies (see also Glover 2015 on rate coefficient uncertainties).
As shown in Table 1, however, we can conclude from our calculations that the impact of Lyman
$\alpha$ feedback can be strong and, importantly, results in a
particularly large drop in the critical LW background flux required
for DCBH formation for background radiation temperatures
characteristic of young, hot stars in the early Universe
(e.g. Tumlinson et al. 2001; Bromm et a. 2001; Oh et al. 2001; Schaerer 2002).

\begin{figure*}
  \includegraphics[angle=270,width=7.2in,trim={6cm 1.5cm 6cm 1.5cm},clip]{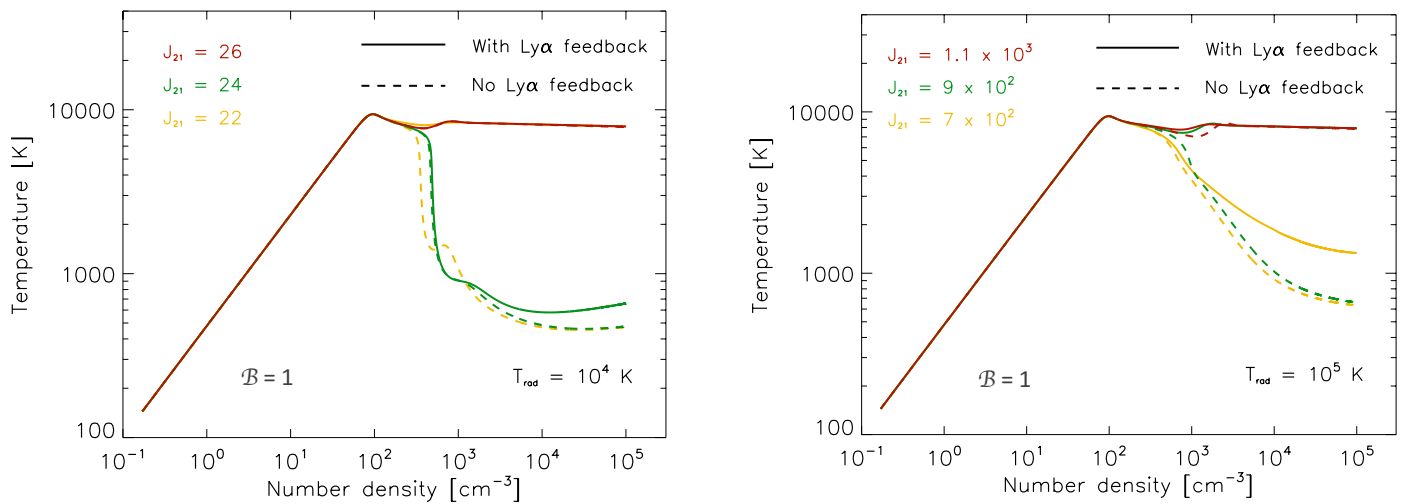}
  \caption{Just as Figure 2, but now with photodetachment of H$^{-}$
    ions by the background radiation field, assumed to be described by
    a blackbody spectrum with a temperature of $T_{\rm rad}$ = 10$^4$
    K ({\it left panel}) and 10$^5$ K ({\it right panel}), included.
    The values of $J_{\rm 21}$ shown in each panel bracket the
    critical values required to maintain the temperature at $\sim$
    10$^4$ K at a density of 10$^5$ cm$^{-3}$ that are inferred both with and without Lyman $\alpha$
    feedback included, as summarized in Table 1.  The case shown here
    assumes $\mathcal{B}$ = 1, corresponding to uniform Lyman $\alpha$ emission
    within the collapsing cloud.}
\end{figure*}

\begin{figure*}
  \includegraphics[angle=270,width=7.2in,trim={6cm 1.5cm 6cm 1.5cm},clip]{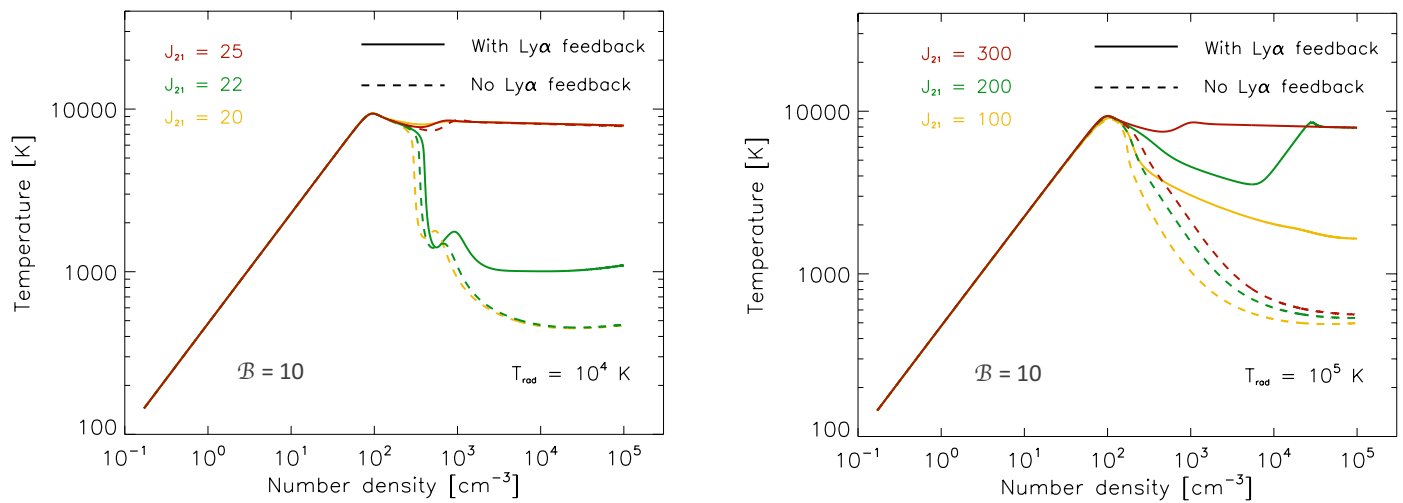}
  \caption{Just as Figure 4, but now with a Lyman $\alpha$ flux
    enhancement $\mathcal{B}$ = 10, an extreme case expected for strongly
    centralized Lyman $\alpha$ emission.}
\end{figure*}

Recent analyses have pointed out that it is more accurate to go beyond $J_{\rm crit}$,
and quantify the requirements for direct collapse in terms of both the
photodetachment rate of H$^-$ and the photodissociation rate of
H$_{\rm 2}$ (e.g. Sugimura et al. 2014; Agarwal et al. 2016a; Wolcott-Green
et al. 2016).  These works show that once $R_{\rm detach} \gsim
10^{-7}$ s$^{-1}$, the photodissociation rate that is required for
direct collapse decreases rapidly. Our calculations indicate that the
constraint $R_{\rm detach} \gsim 10^{-7}$ s$^{-1}$ is reached for
$\log [n/{\rm cm}^{-3}]\gsim$ 2.5, implying that the thermal evolution of the
gas at these high densities becomes strongly impacted by Lyman
$\alpha$ feedback.

\section{Discussion and Conclusions}
We have applied a one-zone chemical and thermal evolution model to
investigate the role that trapped Lyman $\alpha$ cooling radiation, generated
during the collapse of atomic cooling halos, has in suppressing
molecular cooling.  We find that, while this feedback from Lyman
$\alpha$ emission is not strong enough on its own to suppress H$_{\rm
  2}$ cooling, it does have the effect of lowering the intensity of
the background LW radiation level that is required for the
formation of DCBHs.  While our modeling is simplified, the effect
can be pronounced, potentially dropping the critical LW flux by up to
a factor of a few for the background radiation temperatures expected to be
produced by young, hot stars in the early Universe.

One implication of our results is that the number density of DCBHs may
be higher than previously anticipated based on
calculations neglecting H$^-$ detachment by cooling radiation.  Previous works have shown
that the number density of DCBHs increases roughly as $J_{\rm 21,crit}^{-4}$
(Dijkstra et al. 2008, 2014; Inayoshi \& Tanka 2015; Chon et al. 2016),
which suggests that the impact of H$^{-}$ detachment by Lyman $\alpha$
photons results in a large increase of up to a factor of order 10$^2$
in the number density of DCBHs in regions of the early universe
illuminated by bright, young stellar populations.  This is 
important, as DCBH formation may have to occur relatively early in the
epoch of galaxy formation, when stellar populations are still young, in order to be the seeds for the
highest-redshift quasars. The lower values of $J_{\rm 21,crit}$
implied by our results also mean that overall higher rates of DCBH formation may be
realized, perhaps high enough for DCBHs to account for the BHs residing in the
centers of a fraction of normal galaxies today (e.g. Habouzit et
al. 2016).  We do note, however, that perhaps the most likely sources
of the LW radiation that leads to DCBH formation are metal-enriched
stellar populations which are likely to emit radiation with
characteristic temperatures intermediate between the 10$^4$ and 10$^5$
K that we have considered here (e.g. Agarwal et al. 2012; Johnson et
al. 2013).  The precise enhancement of the DCBH formation rate that is
due to Lyman $\alpha$ feedback will clearly depend on the spectra of
the sources producing the LW radiation, and it is possible that if the
spectra are sufficiently soft then the impact of this feedback may be limited.

The extremely bright Lyman $\alpha$ emitter known as CR7 is an
intriguing candidate for a high-redshift quasar that may be powered by
accretion onto a DCBH (Sobral et
al. 2015).  Recent modeling efforts have demonstrated that a nearby
galaxy may well have produced a high enough level of LW
radiation to induce the formation of a DCBH in this galaxy and that
the nebular emission could be explained by an accreting BH with a mass
consistent with formation as a DCBH (e.g. Pallottini et
al. 2015; Agarwal et al. 2016c; Hartwig et al. 2016; Smidt et
al. 2016; Smith et al. 2016; Dijkstra et al. 2016a).\footnote{We note that recent
  observations of CR7 suggest that the bright Lyman $\alpha$ source
  may be enriched to some degree with heavy elements (Bowler et
  al. 2016), suggesting that it is somewhat
  evolved if it did intially host the formation of a DCBH (see
  e.g. Aykutalp et al. 2014; Agarwal et al. 2017).}  In suggesting that the critical LW flux may
be lower than previously thought, our results lend support to DCBH
scenario for the origin of CR7.  We note that this is also consistent
with recent work suggesting that a massive cluster of Population
III stars, an alternative explanation for the origin of
CR7 (e.g. Sobral et al. 2015; Visbal et al. 2016; see also Johnson 2010), is dubious since it
is unknown how a sufficiently high mass of Population III stars
could be assembled rapidly enough to explain the observed extremely
bright emission (e.g. Hartwig et al. 2016; Yajima \& Khochfar 2016; Xu et al. 2016; Visbal et al. 2017).

We note that we have neglected the 2-photon and other hydrogen
line emission that is produced at very high densities ($\ga$ 10$^6$
cm$^{-3}$) where Lyman $\alpha$ photons can be destroyed before
escaping the collapsing cloud (e.g. Schleicher et al. 2010; Dijkstra et al. 2016b).  While
not resonant emission lines, these photons are energetic enough to
detach H$^-$ and, in fact, the cross section for this process is
greater at these photon energies than for Lyman $\alpha$ photons
(e.g. de Jong 1972).  Thus, neglecting this emission may also lead to a
slight overestimate of the critical LW flux.  

We note also, though, that we have neglected
the absorption of Lyman $\alpha$ photons by H$_{\rm
  2}$ molecules, as described in Neufeld (1990; see also Dijkstra et
al. 2016b).  However, we estimate that
this results in a reduction in the Lyman $\alpha$ flux of, at most, a
factor of two at the column densities ($N_{\rm H}$ $\sim$ 10$^{23}$
cm$^{-2}$) and the low H$_{\rm 2}$ fractions ($f_{\rm H2}$ $\sim$ 10$^{-7}$)
that occur with an elevated background radiation field.  In addition,
the LW photons produced in the subsequent radiative decay of the
H$_{\rm 2}$ molecules are also able to detach H$^{-}$.  Thus, we do
not expect that accounting for this effect would strongly impact our
conclusions.  

Our results also carry implications for the impact of X-rays on the
collapse of gas in atomic cooling halos, which numerous authors have shown is to
produce free electrons which catalyze H$_{\rm 2}$ formation, resulting
in an increase in the critical flux $J_{\rm 21,crit}$ (e.g. Inayoshi \& Omukai 2011;
Inayoshi \& Tanaka 2015; Latif et al. 2015; Glover 2016; Regan et
al. 2016b).  We note, in particular, that our results for the critical
LW flux for DCBH formation are in reasonable agreement with those of
Glover (2016) for the case neglecting X-ray feedback.  While X-rays
may have the effect of raising the critical flux by up to two orders
of magnitude in the absence of Lyman $\alpha$ feedback for a hard
spectrum (Glover 2016), another impact of X-rays is to enhance the Lyman $\alpha$
emission within the halo (e.g. Dijkstra et al. 2016a).  As we have
shown, this should result in an enhanced rate of H$^-$ photodetachment 
that will again lower the critical flux.

Finally, we note that atomic cooling halos which grow rapidly, due to
mergers or due to growth in high density environments, likely produce
Lyman $\alpha$ cooling radiation at a higher rate than assumed in our
calculations.  This more intense emission leads, in turn, to larger
photodetachment rates and lower values for the critical
externally-produced LW flux required for DCBH formation.  As the
earliest supermassive black holes form in relatively rare, overdense
regions, this implies that Lyman $\alpha$ feedback may play an
especially strong role in paving the way for the formation of the DCBH
seeds of the earliest bright quasars (e.g. Mortlock et al. 2011; Wu et
al. 2015).  Future work incorporating the feedback effect of Lyman $\alpha$
radiation on the chemical evolution of atomic cooling halos in 3D
cosmological simulations will further elucidate the role that this
process plays in determining the overall rate of DCBH formation.

\section*{Acknowledgements}
Work at LANL was done under the auspices of the National Nuclear Security 
Administration of the US Department of Energy at Los Alamos National 
Laboratory under Contract No. DE-AC52-06NA25396.  JLJ would like to
thank Aaron Smith, Bhaskar Agarwal, Zoltan Haiman, Marta Volonteri,
Aycin Aykutalp, Nicole Lloyd-Ronning and Simon Glover for helpful
discussions.  JLJ also acknowledges support for this work from the
LANL LDRD program.


\appendix

\section{Radial Dependence of Ly$\alpha$ Trapping}\label{app:trap}

In one-zone models, the physical conditions of the collapsing gas
cloud are described completely by its temperature and density. When we
interpret one-zone models as clouds of uniform density in which
Ly$\alpha$ emission is produced uniformly (as we did when deriving
equation (7), the energy density in Ly$\alpha$ photons is enhanced
almost uniformly throughout the cloud (see discussion below and Figure
A.1). The spatial diffusion of Ly$\alpha$ photons out of the cloud introduces only small gradients in the Ly$\alpha$ energy density. However, larger gradients exist if Ly$\alpha$ is not emitted uniformly throughout the cloud, as is generally the case in more realistic scenarios, in which we expect Ly$\alpha$ cooling to increase towards the center of the cloud.

Here we compute the radial dependence of the Ly$\alpha$ energy density
in a suite of spherical gas clouds. We vary the HI column density of
the cloud and where Ly$\alpha$ is emitted, and compute $\mathcal{B}$
by comparing this energy density to our estimate for $u_{\alpha}$
given by equation (4). Ly$\alpha$ transfer through static, spherical
clouds of uniform density can generally be solved analytically for
large line-center optical depths $\tau_0$. Dijkstra et al. (2006)
derive expressions for the (angle-averaged) Ly$\alpha$ intensity $J$
as a function of radius $r$ and frequency $x$ in a spherical cloud of
radius $R_{\rm cl}$ (see their equation C12). For a central Ly$\alpha$ point source (at $r_s=0$) their expression for the total (integrated over frequency) intensity simplifies to

\begin{equation}
J(r/R)=A\sum_{n=1}^{\infty}\int_{-\infty}^{\infty}dx \left(\frac{R}{r}\right) \sin p_n\exp\left(\frac{-\lambda_n|\sigma(x)|}{\kappa_0}\right),
\end{equation} where $A$ is a normalization constant, and
\begin{eqnarray}
p_n&= \pi n \left(\frac{r}{R}\right)\left( 1- \frac{2}{3\tau_0 \phi(x)}\right)\\
\frac{\lambda_n}{\kappa_0}& =\frac{\pi n}{\tau_0}\left( 1- \frac{2}{3\tau_0 \phi(x)}\right)\\
\sigma&=\sqrt{\frac{2\pi}{27}}\frac{x_3}{a_v}.
\end{eqnarray} We obtain $\mathcal{B}$ by dividing the energy density
$u_{\alpha}(r/R)=4\pi J(r/R)/c$ to $u_{\alpha}$ given by equation (4).
\begin{figure}\label{fig:app}
  \includegraphics[angle=270,width=9.0cm]{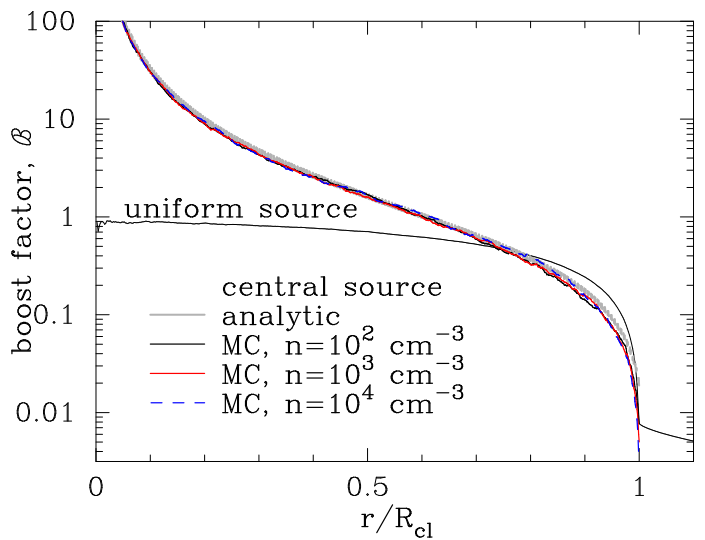}
  \caption{The {\it lines} show the radial dependence of the boost factor $\mathcal{B}$ for a central Ly$\alpha$ point source surrounding a uniform gas cloud of density $n=10^2$ cm$^{-3}$ ({\it black}), $n=10^3$ cm$^{-3}$ ({\it red}), and $n=10^4$ cm$^{-4}$ ({\it blue}). When normalized to the cloud radius $R$, $\mathcal{B}(r/R)$ does not depend on $n$. The {\it grey line} shows analytic solution from Dijkstra et al. (2006). The {\it black line} shows a case in which Ly$\alpha$ is produced uniformly throughout the cloud, and $\mathcal{B}\sim 1$, which shows that the calculations presented in the paper are quite accurate if Ly$\alpha$ is emitted uniformly throughout the cloud.}
\end{figure}

Figure A.1 shows $\mathcal{B}$ as a function of $r/R$ for the analytic
model ({\it thick grey line}). This line is independent of $\tau_0$
provided that $a_v\tau_0 \gsim 10^3$. We overplot 3 lines with
different colors, which we obtained from Monte-Carlo simulations of
the Ly$\alpha$ radiative transfer. The blue, red, and black lines
represent the cloud when its density is $n=10^2$, $10^3$, and 10$^4$
cm$^{-3}$, respectively. First, we note that (not shown here) the
total {\it average} trapping time we found for Ly$\alpha$ photons in
the Monte-Carlo simulation agreed well with our estimate used for
equation (4). Figure A.1 shows clearly that $\mathcal{B}>1$ at $ r< 0.6R_{\rm cl}$, and that $\mathcal{B}>10$ at $r < 0.2 R_{\rm cl}$. That is, in the case of a central Ly$\alpha$ source, photodetachment of $H^-$ by Ly$\alpha$ is signficantly more important for the inner $\sim 10^4 M_{\odot}$ of gas than in the exterior regions. Clearly, the case of a central point source represents an extreme case of centrally enhanced Ly$\alpha$ emission, and we consider the values of $\mathcal{B}$ that we obtain for these models to represent upper limits. 

For completeness, the {\it black line} shows $\mathcal{B}$
obtained from our Monte-Carlo simulations in which Ly$\alpha$ photons
are produced uniformly throughout the cloud. For clarity, we have only
shown the case $n=10^2$ cm$^{-3}$, but we have verified that the curve
does not change for higher densities. The Ly$\alpha$ energy density is
enhanced close to uniformly throughout the cloud, and at a level that
is in good agreement with equation (4).

\end{document}